\documentclass[
  aps,
  prd,
  preprint,
  showpacs,
  nofootinbib
]{revtex4}

\usepackage{amsmath,amssymb}
\usepackage{graphicx}

\DeclareMathOperator{\tr}{tr}

\graphicspath{{./fig/}}

\begin{document}

\preprint{Brown/HET-1585}

\title{The two-loop MHV amplitudes in $\mathcal{N}=4$ supersymmetric Yang-Mills theory}

\author{C.~Vergu}
\email{Cristian_Vergu@brown.edu}
\affiliation{Physics Department, Brown University, Providence, RI 02912, USA}

\begin{abstract}
We compute the even part of the planar two-loop MHV amplitude in $\mathcal{N}=4$ supersymmetric Yang-Mills theory, for an arbitrary number of external particles.  The answer is expressed as a sum of conformal integrals.
\end{abstract}


\maketitle

\section{Introduction}
\label{sec:intro}

In the last few years, scattering amplitudes in $\mathcal{N}=4$ super Yang-Mills theory have been the subject of intense study.  One motivation for this interest is the desire to probe the $\mathcal{N}=4$ theory in different ways than before.

Related to this, different questions can be asked and, indeed, some of them have already been answered.  One natural question is: What is the form of the scattering amplitudes at strong 't Hooft coupling?

The first steps towards answering this question were made in Ref.~\cite{Anastasiou:2003kj} where Anastasiou, Bern, Dixon and Kosower discovered an iteration relation for the two-loop splitting functions and proposed that a similar relation holds for the MHV scattering amplitudes.  Later, in Ref.~\cite{Bern:2005iz}, Bern, Dixon and Smirnov conjectured that a similar relation holds to all orders in perturbation theory, after performing a three-loop check.  This is known as the BDS ansatz in the literature.

By the AdS/CFT correspondence (see Refs.~\cite{Maldacena:1997re, Gubser:1998bc, Witten:1998qj}) we know that the natural framework to address questions about the strongly-coupled gauge theory is to use the gravitational dual of the theory.  The scattering amplitudes have been analyzed in this context in a ground-breaking paper by Alday and Maldacena (see Ref.~\cite{Alday:2007hr}).  This paper hinted that there may be a connection between scattering amplitudes and Wilson loops in $\mathcal{N}=4$ super Yang-Mills.

The connection of the scattering amplitudes with the Wilson loops was known to hold for the infrared divergent parts.  (More precisely, the IR divergences of the scattering amplitude can be put into correspondence with the UV divergences of some Wilson loop with cusps. See Refs.~\cite{Sudakov:1954sw, Sen:1981sd, Ivanov:1985np, Collins:1989bt, Magnea:1990zb, Korchemskaya:1992je, Korchemsky:1992xv, Catani:1998bh, Sterman:2002qn} for more information on IR divergences of the scattering amplitudes.)

The connection between Wilson loops and MHV scattering amplitudes was confirmed by weak coupling computations in Refs.~\cite{Drummond:2007aua, Brandhuber:2007yx, Drummond:2007cf, Drummond:2007au, Anastasiou:2009kn} for Wilson loops.  See also Refs.~\cite{Bern:2006vw, Cachazo:2006tj, Bern:2007ct, Bern:2008ap, Cachazo:2008hp, Spradlin:2008uu, Vergu:2009zm} for scattering amplitudes computations.  At strong coupling, the connection between Wilson loops and scattering amplitudes was analyzed in Refs.~\cite{Ricci:2007eq, Beisert:2008iq, Berkovits:2008ic}.

Unfortunately the BDS ansatz was shown to break down (see Refs.~\cite{Alday:2007he, Drummond:2007bm, Drummond:2008aq, Bern:2008ap}).  Correcting the BDS ansatz is still an important open problem.  Recently there has been some progress in finding the strong-coupling answers for higher-point scattering amplitudes (see Refs.~\cite{Alday:2009ga, Alday:2009yn}).

Another question about scattering amplitudes is: What is the link with integrability?  Indeed, it is known that the $\mathcal{N}=4$ theory is integrable (at strong coupling the worldsheet sigma model describing the planar limit of the dual string theory is integrable as shown in Ref.~\cite{Bena:2003wd}, while at weak coupling the dilatation operator is the Hamiltonian of an integrable system, as shown in Ref.~\cite{Minahan:2002ve} and subsequent papers).

It is fair to say that the implications of integrability for the scattering amplitudes are not completely understood.  In this context, it is worthwhile to mention that a new symmetry was discovered, the dual conformal symmetry.  This symmetry was proposed and studied in a series of papers~\cite{Drummond:2006rz, Drummond:2007aua, Drummond:2007au, Drummond:2008bq, Brandhuber:2009xz, Elvang:2009ya, Hall:2009xg} and it was later extended to a dual conformal supersymmetry in Ref.~\cite{Drummond:2008vq}.  It was later understood (see Ref.~\cite{Drummond:2009fd}) that the dual conformal supersymmetry is part of a Yangian structure.

It is fascinating to ask if the amplitudes are completely constrained by the symmetries.  This question has been investigated recently in Refs.~\cite{Bargheer:2009qu, Korchemsky:2009hm}.  However, a more detailed understanding is needed at loop level, since some of these symmetries are broken by quantum corrections.

A different line of attack in understanding the scattering amplitudes is by studying them in twistor space.  The twistor space approach was pioneered by Penrose (see Refs.~\cite{Penrose:1967wn, Penrose:1972ia}) and was revived more recently following work by Witten (see Ref.~\cite{Witten:2003nn}).

The twistor constructions are still mysterious and poorly understood, but there are strong hints that the twistor picture can be very useful.  Indeed, some recent papers (see Refs.~\cite{ArkaniHamed:2009si, ArkaniHamed:2009dn}) have been re-emphasizing the view that space-time physics should play a secondary role, and all the computations should be done in twistor language.  This is in line with what Penrose proposed in his early work on twistor theory.

However, one would like to be able to relate the ``space-time'' physics to the ``dual'' twistor space picture, and so far this has proven to be an elusive target.  Despite the difficulty of the problem, some progress has been made.  The BCFW\footnote{The supersymmetric version of these recursion relations first discussed in Refs.~\cite{Brandhuber:2008pf, ArkaniHamed:2008gz} are naturally represented in the supersymmetric version of the ambi-twistor space.} recursion relations (see Ref.~\cite{Britto:2005fq}) were given a natural translation in ambi-twistor space in Refs.~\cite{ArkaniHamed:2009si, Mason:2009sa} where they were shown to be related to the diagrammatic approach of Hodges (see Ref.~\cite{Hodges:2005bf}).

The action of the conformal symmetry is obvious in twistor space, but the dual conformal symmetry is not.  In a recent paper~\cite{Hodges:2009hk}, Hodges introduced the ``momentum-twistors,'' which are obtained by a twistor transform of the dual space, which is the space on which the dual conformal transformations act.  Interestingly, in this language the full dihedral symmetry of the planar amplitudes becomes manifest.

\section{Review}
\label{sec:review}

The $n$-point $\ell$-loop planar $\mathcal{N}=4$ scattering amplitude can be written as follows
\begin{equation}
  \label{eq:color_decomposition}
  \mathcal{A}_n^{(\ell)} = g^{n-2} a^\ell \sum_{\sigma \in \mathcal{S}_n/\mathbb{Z}_n} \tr \left( T^{\sigma(1)} \cdots T^{\sigma(n)}\right) A_n^{(\ell)}(\sigma(1), \dotsc, \sigma(n)),
\end{equation} where $T^a$ are the generators of the $SU(N_c)$ gauge group and the sum runs over the cyclically unrelated permutations of the external lines.  The $su(N_c)$ algebra generators are normalized by $\tr (T^a T^b) = \delta^{a b}$ (which is different from the usual normalization $\tr (T^a T^b) = \tfrac{1}{2} \delta^{a b}$).  We have also used the notation
\begin{equation}
  \label{eq:a_def}
  a \equiv (4 \pi e^{-\gamma})^\epsilon \frac {\lambda}{8 \pi^2},
\end{equation} where $\epsilon = \frac{4-D}{2}$ is the dimensional regularization parameter, $\lambda = g^2 N_c$ is the 't Hooft coupling and $\gamma = - \Gamma'(1)$ is the Euler constant.

For MHV amplitudes the supersymmetry insures that the loop amplitudes are proportional to the tree amplitudes and therefore it is convenient to define the ratios
\begin{equation}
  \label{eq:m_def}
  M_n^{(\ell)} = \frac {A_n^{(\ell)}}{A_n^{(0)}}.
\end{equation}  The quantities $M_n$ are scalars and can be written as a sum of a scalar and a pseudoscalar part (also called parity even and parity odd parts).  In the MHV case, the quantities $M_n^{(\ell)}$ can be decomposed over a basis of integrals.

In a Feynman diagram computation the integrals are normalized as
\begin{equation}
  \label{eq:nomalization1}
  \int \frac {d^D p}{(2 \pi)^D},
\end{equation} but by taking into account the normalization in Eq.~\eqref{eq:color_decomposition} we get an extra factor of $(4 \pi)^{\frac{D}{2}} e^{\gamma \epsilon}$ per loop.  In fact, it is conventional to also include a factor of $-i$ to obtain a normalization by
\begin{equation}
  \label{eq:nomalization2}
  - i \pi^{-\frac{D}{2}} e^{\gamma \epsilon} \int d^D p.
\end{equation}  After normalizing the integrals in this way, the only loop dependent factor in Eq.~\eqref{eq:color_decomposition} is $\left(\tfrac {\lambda}{2}\right)^\ell$.

The integrals appearing in the result of a unitarity computation can be thought of as arising from a corresponding Feynman diagram computation after applying a large number of algebraic transformations.  Even though in practice these algebraic transformations can generally not be applied explicitly because of an explosion in complexity, they can be applied in principle.  This is important because the transformations preserve the momentum conservation constraints so the integrals after reduction have the same structure as integrals arising from Feynman diagrams.  It is therefore appropriate to associate a graph to each of these integrals.  
This associated graph doesn't have any degree two vertices\footnote{The \emph{degree} of a vertex is the number of incident edges.} and the only degree one vertices are the external vertices (of course, when we write down the integral associated to a given diagram we don't include propagators for the external on-shell momenta).  This restriction on the degree of the vertices is obvious because the algebraic transformations can only increase the degree of a vertex (by eliminating an incident edge or, in Feynman diagram language, by canceling a propagator) and the minimal degree is three.

Note also the we can restrict to 1PI graphs because each one-particle reducible graph differs from the 1PI graph obtained by collapsing all the bridges\footnote{A \emph{bridge} is an edge whose removal disconnects the graph.} by a multiplicative factor determined uniquely by the external momenta.

From now on, when we talk about graphs we mean planar graphs subject to the above restrictions. 

\section{Description of the computational procedure}
\label{sec:description}

The computation was done by using the unitarity method (see Refs.~\cite{Bern:1994zx, Bern:1994cg, Bern:1997sc}).  For applying this method, we need to find a set of cuts which detect all the possible integrals appearing in the result.

The graphs associated to the integrals (see Sec.~\ref{sec:review} for a discussion) can be divided in two categories: the graphs where the two loops share an edge, and the graphs where the two loops only share a vertex\footnote{This is called a \emph{cut vertex} or an \emph{articulation point} in graph-theory language.} (the latter have been called ``kissing topologies'').  Then, the external lines can be attached to the left or right loops or they can be attached to the vertices incident with the edge common to both loops or to the vertex shared by the the two loops.  Because we are computing a color ordered amplitude the ordering of external legs is important.

\begin{figure}
  \centering
  \includegraphics{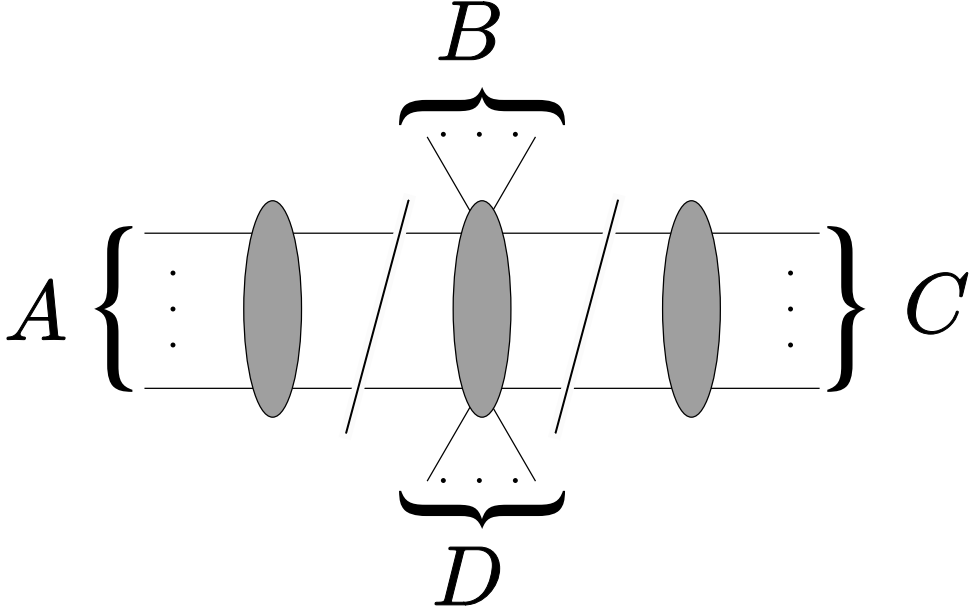}
  \caption{The generic form of the cuts we computed.  The subsets of particles $B$ and $D$ can be empty while the number of particles in the subsets $A$ and $C$ is at least two.}
  \label{fig:cuts}
\end{figure}

It should be clear that all the graphs satisfying the conditions formulated above can be cut as in Fig.~\ref{fig:cuts}, for some distribution of the external legs in the subsets $A$, $B$, $C$ and $D$.  Below we will use two-particle cuts of the type described in Fig.~\ref{fig:cuts}.  Some other types of cuts are possible, but the ones we will use have the advantage that they can be computed using uniquely MHV tree-level amplitudes.

In practice, one should distinguish between several cases.  The differences between these cases arise from the different form of the tree amplitudes and of the momentum conservation conditions.  If the sets $A$, $B$, $C$ and $D$ contain more that two external lines then, after dividing by the tree-level amplitude, the answer will only depend on the first and the last particle in each set (here we order the sets in the same way as the external particles).  Moreover, it is easy to see that there are no constraints among the momenta which are present in the ratio to the tree amplitude, because the momenta of the ``missing'' particles can be adjusted to satisfy momentum conservation.  This is not the case if the sets $A$ or $C$ contain only two particles.  Then, there are momentum conservation constraints that must be taken into account and this will change the form of the result.  This necessitates a special treatment for these cases.  Other special cases occur when either of the sets $B$ and $D$ has zero, one or two particles.

Considering all these possibilities, one can see that the cuts in Table~\ref{tab:cuts} are sufficient to detect all the integrals, or their symmetric under the group generated by the horizontal and vertical flips.  Note that some of these cuts are specific to some fixed number of particles and have been computed before.  Two cuts for an arbitrary number of particles have been computed in Ref.~\cite{Vergu:2009zm}.

Our method of computation, which we will describe in more detail below, requires that the answer be expressible in terms of a finite number of types of integrals, even though the number of external legs is unbounded.  For this reason we will impose a number of restrictions on the integrals we include in the ansatz.  Our computation will confirm that the ansatz has all the right singularities.

Unlike in the one-loop case, at two loops no basis of integrals is known but, based on previous computations, one can make some reasonable assumptions.  For example, we don't expect to find any integrals with a triangle sub-diagram.  We also expect to find only conformal integrals in the computation of the even part (even though this expectation has not been proven, it has been borne out by explicit computations).  Note that the assumption that only conformal integrals appear in the even part of the amplitude implies the absence of integrals with triangle and bubble sub-diagrams.

\begin{table}
  \centering
  \label{tab:cuts}
  \begin{tabular}{| c | c | c | c | c | c |}
    \hline
    nr & $A$ & $B$ & $C$ & $D$ & $n$\\
    \hline
    $1$ & $2$ & $0$ & $2$ & $0$ & $n = 4$\\
    $2$ & $2$ & $0$ & $2$ & $1$ & $n = 5$\\
    $3$ & $2$ & $0$ & $2$ & $2$ & $n = 6$\\
    $4*$ & $2$ & $0$ & $2$ & $>2$ & $n \ge 7$\\
    $5$ & $2$ & $0$ & $>2$ & $0$ & $n \ge 5$\\
    $6*$ & $2$ & $0$ & $>2$ & $1$ & $n \ge 6$\\
    $7*$ & $2$ & $0$ & $>2$ & $2$ & $n \ge 7$\\
    $8*$ & $2$ & $0$ & $>2$ & $>2$ & $n \ge 8$\\
    $9$ & $2$ & $1$ & $2$ & $1$ & $n = 6$\\
    $10$ & $2$ & $1$ & $2$ & $2$ & $n = 7$\\
    $11*$ & $2$ & $1$ & $2$ & $>2$ & $n \ge 8$\\
    $12*$ & $2$ & $1$ & $>2$ & $1$ & $n \ge 7$\\
    $13*$ & $2$ & $1$ & $>2$ & $2$ & $n \ge 8$\\
    $14*$ & $2$ & $1$ & $>2$ & $>2$ & $n \ge 9$\\
    $15*$ & $2$ & $2$ & $2$ & $2$ & $n = 8$\\
    \hline
  \end{tabular}
  \hspace{2em}
  \begin{tabular}{| c | c | c | c | c | c |}
    \hline
    nr & $A$ & $B$ & $C$ & $D$ & $n$\\
    \hline
    $16*$ & $2$ & $2$ & $2$ & $>2$ & $n \ge 9$\\
    $17*$ & $2$ & $2$ & $>2$ & $2$ & $n \ge 9$\\
    $18*$ & $2$ & $2$ & $>2$ & $>2$ & $n \ge 10$\\
    $19*$ & $2$ & $>2$ & $2$ & $>2$ & $n \ge 10$\\
    $20*$ & $2$ & $>2$ & $>2$ & $>2$ & $n \ge 11$\\
    $21$ & $>2$ & $0$ & $>2$ & $0$ & $n \ge 6$\\
    $22*$ & $>2$ & $0$ & $>2$ & $1$ & $n \ge 7$\\
    $23*$ & $>2$ & $0$ & $>2$ & $2$ & $n \ge 8$\\
    $24*$ & $>2$ & $0$ & $>2$ & $>2$ & $n \ge 9$\\
    $25*$ & $>2$ & $1$ & $>2$ & $1$ & $n \ge 8$\\
    $26*$ & $>2$ & $1$ & $>2$ & $2$ & $n \ge 9$\\
    $27*$ & $>2$ & $1$ & $>2$ & $>2$ & $n \ge 10$\\
    $28*$ & $>2$ & $2$ & $>2$ & $2$ & $n \ge 10$\\
    $29*$ & $>2$ & $2$ & $>2$ & $>2$ & $n \ge 11$\\
    $30*$ & $>2$ & $>2$ & $>2$ & $>2$ & $n \ge 12$\\
    \hline
  \end{tabular}
  \caption{The set of cuts used to compute the planar two-loop MHV amplitude.  The columns two through four list the number of particles in the subsets $A$, $B$, $C$ and $D$ as described in Fig.~\ref{fig:cuts}.  Note that we treat separately the cases with $2$ and $>2$ particles in subsets $A$ and $C$ and also the cases with $0$, $1$, $2$ and $>2$ particles in subsets $B$ and $D$ (the reason for distinguishing between these cases will be explained in the text).  We marked the cuts that we computed in this paper by a star; the unstarred cuts were computed previously.}
\end{table}

Another important restriction which can be proposed on the basis of the results in Ref.~\cite{Vergu:2009zm} is the absence of integrals containing higher polygons.  More precisely, the restriction can be formulated as follows.  Consider the graphs where the two loops share an edge, and the graphs where the two loops only share a vertex.  For the graphs of the first kind we impose the restriction that they should not contain any hexagons or higher polygons, while for the graphs of the second kind we impose the condition that they should not contain any pentagons or higher polygons.  This has been shown to be true in the $n=7$ case by the computation in Ref.~\cite{Vergu:2009zm}, but we consider that evidence to be strong enough to justify its extension beyond $n=7$.  Another way to formulate this constraint is to say that we do not include in our ansatz any integral which has a loop with four or more independent external momenta.  In this formulation, the constraint on the admissible integrals may hold at higher loops and for non-MHV amplitudes.

The constraints one can impose on the integrals are not only of topological nature, as above.  In fact, even after imposing the restrictions above there are many possibilities of attaching the external particles to the underlying topology.  When several massless external particles attach to the same point, they can be effectively treated as a massive external momentum.  A first guess would be that the integral coefficients don't depend on the number of particles attached at each point, but only on whether the legs are massless or massive.  This is supported by the one-loop computation in Ref.~\cite{Bern:1994zx} and also by the leg addition rule of Ref.~\cite{Vergu:2009zm}.

\begin{figure}
  \centering
  \includegraphics[width=.6\textwidth]{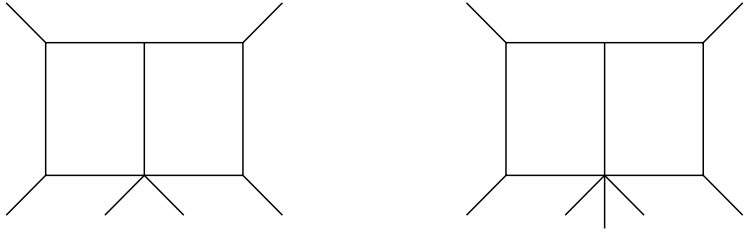}
  \caption{Two integrals with one massive leg, but with different integral coefficients.  The six-point integral has coefficient zero as shown in Ref.~\cite{Bern:2008ap} while the seven-point integral has coefficient non-zero as shown in Ref.~\cite{Vergu:2009zm}.}
  \label{fig:massive}
\end{figure}

However, the results of the six- and seven-point computations suggest that things are not so simple.  Indeed, the six-point integral in Fig.~\ref{fig:massive} was shown to have coefficient zero in Ref.~\cite{Bern:2008ap}, while the seven-point integral in Fig.~\ref{fig:massive} was shown to have a non-zero coefficient in Ref.~\cite{Vergu:2009zm}.

This difference can be explained in the same way as above; when any of the sets $B$ and $D$ of a given cut have two particles the momentum conservation constraint is different, so it's not surprising that some integral coefficients detected by that cut are modified.

\begin{figure}
  \centering
  \includegraphics[width=.8\textwidth]{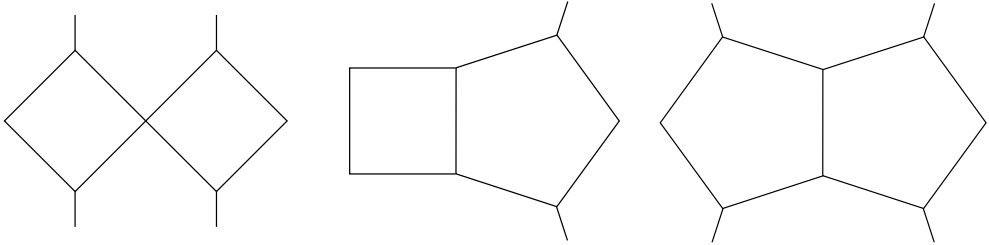}
  \caption{The legs shown in this figure must be massless.  Integrals where at least one of these legs is massive have coefficient zero and therefore don't appear in the MHV result.}
  \label{fig:massless}
\end{figure}

The results of Ref.~\cite{Vergu:2009zm} suggest even further constraints on the positions of the massless and massive external legs.  In the seven-point result the integrals with the topologies in Fig.~\ref{fig:massless} have massless momenta attached at the positions indicated by the external lines.  We will start with the assumption that this holds for an arbitrary number of points.  This assumption is significant because it drastically reduces the number of integrals we need to consider.

Let us recapitulate the ingredients of our ansatz.  We start with a set of integrals with topologies of ``kissing'' double boxes, double boxes, box-pentagon and double pentagons with massless legs in the positions indicated in Fig.~\ref{fig:massless}.  We distinguish between the integrals with two or more legs attached at the vertices of the common edge to the two loops (see Fig.~\ref{fig:massive}).

One can now list all the topologies compatible with the above constraints, and check that there are only a finite number of possibilities.  Each of these topologies can be made into a conformal integral by adding appropriate numerator factors.  For the even part, these numerator factors are conformal factors times rational coefficients.  Our purpose will be to determine these rational coefficients.

As we already mentioned it is essential that only a finite number of unknowns need to be found.  These unknowns can be determined as follows:
\begin{itemize}
\item for each cut in Table~\ref{tab:cuts} find the topologies possessing the required propagators.
\item for each of these topologies find all the possible ways of adding conformal numerators and associate to each of these ``conformal dressings'' an unknown which needs to be determined.  This unknown is the coefficient of the conformal dressing under consideration.
\item compute the product of tree amplitudes and divide by the global tree amplitude.  This should be equal to the sum of integrals, subject to the on-shell conditions imposed by the cuts.  In order to isolate the even part, take sum between this ratio and its parity conjugate\footnote{The parity conjugate of an expression containing spinor products is obtained by swapping angle and square products ($\langle\; \rangle \leftrightarrow [\; ]$).} and divide by two.
\item generate random kinematics satisfying momentum conservation and on-shell constraints.\footnote{It is important to note that the ratio of the product of tree amplitudes to the global tree amplitude only depends on a finite (and rather small) number of momenta.  This circumstance allows a numerical evaluation which otherwise would not be possible.}  By using this kinematics, one can obtain numerical equations for the unknown coefficients.  By generating enough equations and solving the corresponding linear system all the unknown coefficients can be found.  Up to rounding errors all the coefficients are found to be simple rational numbers.
\end{itemize}

Some of the topologies we consider have symmetries which can be used to reduce the number of unknowns.  We have found that using these symmetries greatly reduces the complexity of the computations.  It has also been useful to use the results for the cuts already computed when computing the coefficients of integrals detected by new cuts.  In this way, one can insure the consistency of results from different cuts and reduce the number of unknowns at the same time.

Let us make a further comment about our computation procedure.  The conformal integrals with pentagon loops have numerators containing the loop momenta in combinations like $(k + l)^2$, where $l$ is the loop momentum and $k$ is an external on-shell momentum.  If the propagator with momentum $l$ is cut then, on that cut, one cannot distinguish between $(k + l)^2$ and $2 k \cdot l$.  However, it is easy to see that one can choose to cut another propagator and in that case this ambiguity does not arise and the numerator factor is uniquely defined.

\section{Results}
\label{sec:results}

We use dual variable notation (see Ref.~\cite{Drummond:2006rz}) for the integrals.  The external dual variables are listed in clockwise direction.  To the left loop we associate the dual variable $x_p$ and to the right loop we associate the dual variable $x_q$.  We use the notation $x_{i j} \equiv x_i - x_j$.

We introduce the following notation which will be useful in the following
\begin{equation}
  \label{eq:square-braket-notation}
  \begin{bmatrix}
    a & b & c & \cdots\\
    a' & b' & c' & \cdots
  \end{bmatrix} = x_{a a'}^2 x_{b b'}^2 x_{c c'}^2 \cdots \pm (\text{permutations of $\lbrace a', b', c', \ldots \rbrace$}).
\end{equation}  The sign $\pm$ above takes into account the signature of the permutation of $\lbrace a', b', c', \ldots \rbrace$.  It is easy to show that
\begin{equation}
  \begin{bmatrix}
    a & b & c & \cdots\\
    a' & b' & c' & \cdots
  \end{bmatrix} = \det_{\substack{i \in \lbrace a, b, c, \cdots\rbrace \\ j \in \lbrace a', b', c', \cdots\rbrace}} x_{ij}^2.
\end{equation}

For some topologies, the expansion of the $\begin{bmatrix} & \end{bmatrix}$ symbol yields terms that would cancel propagators.  For those cases we make the convention that all the terms that would cancel propagators are absent.  In fact, as we will see, terms that would cancel propagators of the double pentagon topologies naturally yield coefficients for some of the topologies with a smaller number of propagators.

\subsection{Double box topologies}

In the case of the double box topologies the massive legs attached to the vertices incident with the common edge have to be a sum of at least three massless momenta.  The cases where these massive legs are the sum of two massless momenta are treated separately in the subsection.~\ref{sssec:extra_double}.  This distinction only arises for the double box topologies.

\subsubsection{No legs attached}

\begin{minipage}[c]{0.3\textwidth}
  \includegraphics{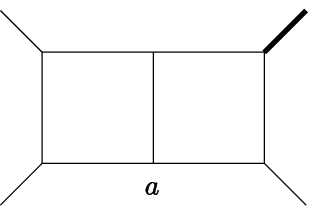}
\end{minipage}
\begin{minipage}[c]{0.7\textwidth}
\begin{equation}
  \label{eq:DBoozmoz}
  \frac{1}{2} \left(x_{a,a+2}^2\right)^2 x_{a-1,a+1}^2
\end{equation}
\end{minipage}

\begin{minipage}[c]{0.3\textwidth}
  \includegraphics{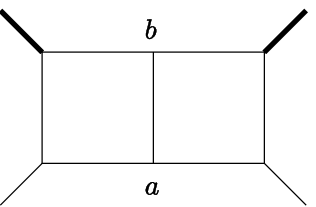}
\end{minipage}
\begin{minipage}[c]{0.7\textwidth}
\begin{equation}
  \label{eq:DBomzmoz}
  \frac{1}{4} \left(x_{a b}^2\right)^2 x_{a-1,a+1}^2
\end{equation}
\end{minipage}

\begin{minipage}[c]{0.3\textwidth}
  \includegraphics{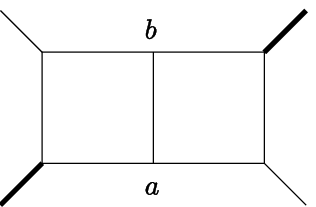}
\end{minipage}
\begin{minipage}[c]{0.7\textwidth}
\begin{equation}
  \label{eq:DBmozmoz}
  -\frac{1}{4} x_{a b}^2 \left(x_{a,b-1}^2 x_{a-1,b}^2 - x_{a b}^2 x_{a-1,b-1}^2\right)^2
\end{equation}
\end{minipage}

\subsubsection{One massless leg attached}

\begin{minipage}[c]{0.3\textwidth}
  \includegraphics{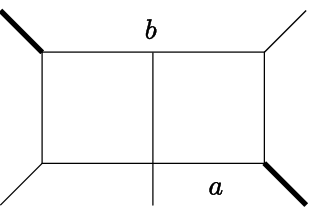}
\end{minipage}
\begin{minipage}[c]{0.7\textwidth}
\begin{equation}
  \label{eq:DBomzomo}
  \frac{1}{4} \left(x_{a,b+1}^2 x_{a+1,b}^2-x_{ab}^2 x_{a+1,b+1}^2\right) x_{a+2,b}^2
\end{equation}
\end{minipage}

\begin{minipage}[c]{0.3\textwidth}
  \includegraphics{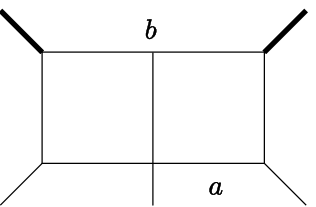}
\end{minipage}
\begin{minipage}[c]{0.7\textwidth}
\begin{multline}
  \label{eq:DBomzmoo}
  \frac{1}{4} \Bigl(-x_{a-1,b}^2 x_{a,a+2}^2 x_{a+1,b}^2+x_{a-1,a+2}^2 x_{a b}^2   x_{a+1,b}^2-\\-x_{a-1,a+1}^2 x_{a b}^2 x_{a+2,b}^2\Bigr)
\end{multline}
\end{minipage}

\begin{minipage}[c]{0.3\textwidth}
  \includegraphics{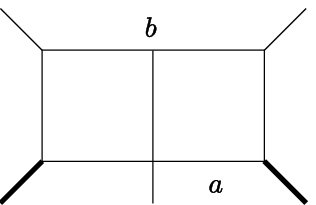}
\end{minipage}
\begin{minipage}[c]{0.7\textwidth}
\begin{equation}
  \label{eq:DBmozomo}
  -\frac{1}{4} x_{ab}^2 x_{a+1,b}^2 x_{b-1,b+1}^2
\end{equation}
\end{minipage}

\begin{minipage}[c]{0.3\textwidth}
  \includegraphics{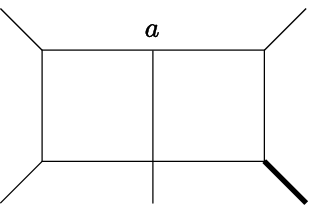}
\end{minipage}
\begin{minipage}[c]{0.7\textwidth}
\begin{equation}
  \label{eq:DBoozomo}
  -\frac{1}{4} x_{a-3,a}^2 x_{a-2,a}^2 x_{a-1,a+1}^2
\end{equation}
\end{minipage}

\begin{minipage}[c]{0.3\textwidth}
  \includegraphics{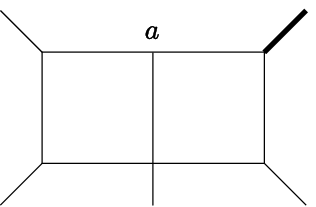}
\end{minipage}
\begin{minipage}[c]{0.7\textwidth}
\begin{equation}
  \label{eq:DBoozmoo}
  \frac{1}{4} \left(x_{a-4,a-1}^2 x_{a-3,a}^2-2 x_{a-4,a}^2 x_{a-3,a-1}^2\right) x_{a-2,a}^2
\end{equation}
\end{minipage}

\subsubsection{Two massless legs attached}

\begin{minipage}[c]{0.3\textwidth}
  \includegraphics{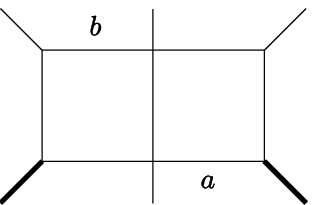}
\end{minipage}
\begin{minipage}[c]{0.7\textwidth}
\begin{multline}
  \label{eq:DBmooomo}
   \frac{1}{4} \Bigl(x_{a,b+2}^2 x_{a+1,b}^2 x_{b-1,b+1}^2-x_{a,b+1}^2 x_{a+1,b}^2 x_{b-1,b+2}^2+\\+x_{a,b+1}^2 x_{a+1,b-1}^2 x_{b,b+2}^2\Bigr)
\end{multline}
\end{minipage}

\begin{minipage}[c]{0.3\textwidth}
  \includegraphics{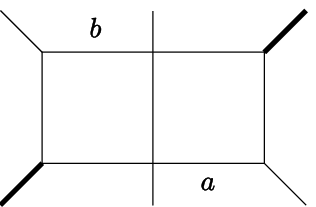}
\end{minipage}
\begin{minipage}[c]{0.7\textwidth}
\begin{multline}
  \label{eq:DBmoomoo}
  \frac{1}{4} \Bigl(-x_{a-1,b-1}^2 x_{a,b+1}^2 x_{a+1,b}^2+x_{a-1,b-1}^2 x_{a b}^2 x_{a+1,b+1}^2-\\-x_{a-1,a+1}^2 x_{a b}^2 x_{b-1,b+1}^2\Bigr)
\end{multline}
\end{minipage}

\begin{minipage}[c]{0.3\textwidth}
  \includegraphics{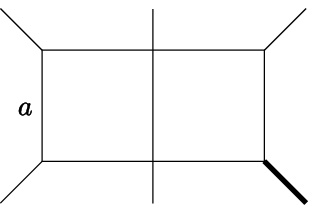}
\end{minipage}
\begin{minipage}[c]{0.7\textwidth}
\begin{multline}
  \label{eq:DBoooomo}
  \frac{1}{4} \Bigl(x_{a-2,a+3}^2 x_{a-1,a+1}^2 x_{a,a+2}^2-2 x_{a-2,a+2}^2 x_{a-1,a+1}^2 x_{a,a+3}^2+\\+x_{a-2,a+1}^2 x_{a-1,a+2}^2 x_{a,a+3}^2-x_{a-2,a}^2 x_{a-1,a+2}^2 x_{a+1,a+3}^2\Bigr)
\end{multline}
\end{minipage}

\subsubsection{One massive leg attached}

\begin{minipage}[c]{0.3\textwidth}
  \includegraphics{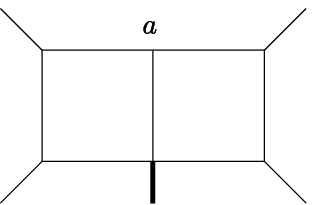}
\end{minipage}
\begin{minipage}[c]{0.7\textwidth}
\begin{equation}
  \label{eq:DBoozoom}
  \frac{1}{4} x_{a-2,a}^2 x_{a-1,a+1}^2 x_{a,a+2}^2
\end{equation}
\end{minipage}

\begin{minipage}[c]{0.3\textwidth}
  \includegraphics{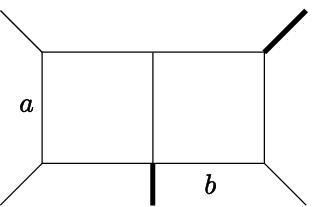}
\end{minipage}
\begin{minipage}[c]{0.7\textwidth}
\begin{equation}
  \label{eq:DBoozmom}
  \frac{1}{4} \left(x_{a-1,a+1}^2 x_{a,b-1}^2 x_{a+1,b}^2-x_{a-1,a+1}^2 x_{ab}^2 x_{a+1,b-1}^2\right)
\end{equation}
\end{minipage}

\begin{minipage}[c]{0.3\textwidth}
  \includegraphics{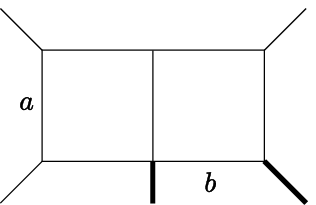}
\end{minipage}
\begin{minipage}[c]{0.7\textwidth}
\begin{equation}
  \label{eq:DBoozomm}
  0
\end{equation}
\end{minipage}

\begin{minipage}[c]{0.3\textwidth}
  \includegraphics{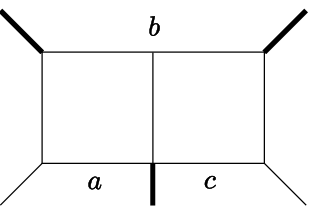}
\end{minipage}
\begin{minipage}[c]{0.7\textwidth}
\begin{multline}
  \label{eq:DBomzmom}
  \frac{1}{4} \Bigl(x_{a c}^2 x_{a+1,b}^2 x_{b,c-1}^2-x_{a b}^2 x_{a+1,c}^2 x_{b,c-1}^2-\\-x_{a,c-1}^2 x_{a+1,b}^2 x_{b c}^2+x_{a b}^2 x_{a+1,c-1}^2 x_{b c}^2\Bigr)
\end{multline}
\end{minipage}

\begin{minipage}[c]{0.3\textwidth}
  \includegraphics{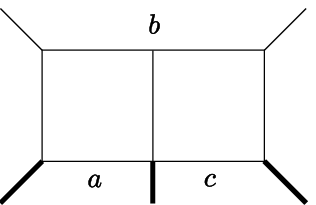}
\end{minipage}
\begin{minipage}[c]{0.7\textwidth}
\begin{equation}
  \label{eq:DBmozomm}
  0
\end{equation}
\end{minipage}

\begin{minipage}[c]{0.3\textwidth}
  \includegraphics{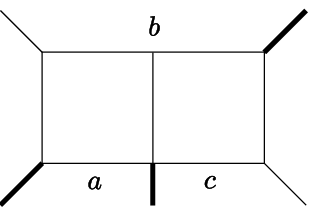}
\end{minipage}
\begin{minipage}[c]{0.7\textwidth}
\begin{equation}
  \label{eq:DBmozmom}
  0
\end{equation}
\end{minipage}

\subsubsection{One massless leg and one massive leg attached}

\begin{minipage}[c]{0.3\textwidth}
  \includegraphics{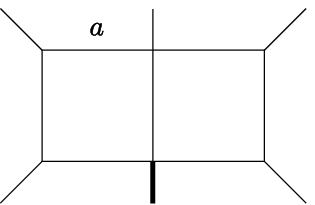}
\end{minipage}
\begin{minipage}[c]{0.7\textwidth}
\begin{equation}
  \label{eq:DBooooom}
  -\frac{1}{4} x_{a-2,a}^2 x_{a-1,a+2}^2 x_{a+1,a+3}^2
\end{equation}
\end{minipage}

\begin{minipage}[c]{0.3\textwidth}
  \includegraphics{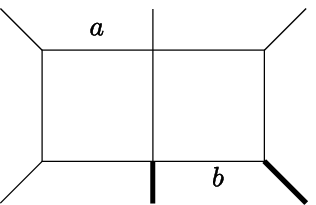}
\end{minipage}
\begin{minipage}[c]{0.7\textwidth}
\begin{equation}
  \label{eq:DBoooomm}
  0
\end{equation}
\end{minipage}

\begin{minipage}[c]{0.3\textwidth}
  \includegraphics{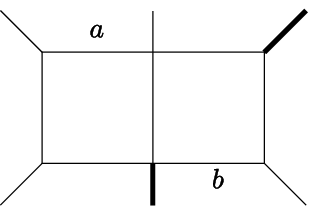}
\end{minipage}
\begin{minipage}[c]{0.7\textwidth}
\begin{equation}
  \label{eq:DBooomom}
  \frac{1}{4} x_{a-2,a}^2 \left(x_{a-1,b}^2 x_{a+1,b-1}^2-x_{a-1,b-1}^2 x_{a+1,b}^2\right)
\end{equation}
\end{minipage}

\begin{minipage}[c]{0.3\textwidth}
  \includegraphics{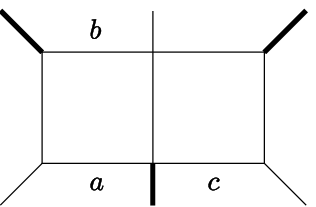}
\end{minipage}
\begin{minipage}[c]{0.7\textwidth}
\begin{multline}
  \label{eq:DBomomom}
  \frac{1}{4} \Bigl(-x_{a c}^2 x_{a+1,b}^2 x_{b+1,c-1}^2+x_{a b}^2 x_{a+1,c}^2 x_{b+1,c-1}^2+\\+x_{a,c-1}^2 x_{a+1,b}^2 x_{b+1,c}^2-x_{a b}^2 x_{a+1,c-1}^2 x_{b+1,c}^2\Bigr)
\end{multline}
\end{minipage}

\begin{minipage}[c]{0.3\textwidth}
  \includegraphics{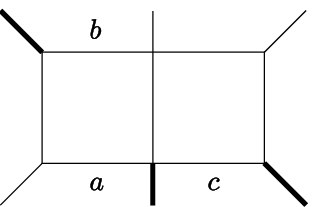}
\end{minipage}
\begin{minipage}[c]{0.7\textwidth}
\begin{equation}
  \label{eq:DBomoomm}
  0
\end{equation}
\end{minipage}

\begin{minipage}[c]{0.3\textwidth}
  \includegraphics{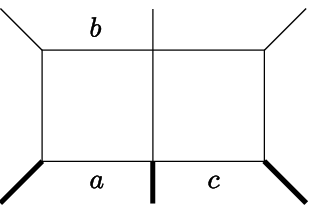}
\end{minipage}
\begin{minipage}[c]{0.7\textwidth}
\begin{equation}
  \label{eq:DBmooomm}
  0
\end{equation}
\end{minipage}

\subsubsection{Two massive legs attached}

\begin{minipage}[c]{0.3\textwidth}
  \includegraphics{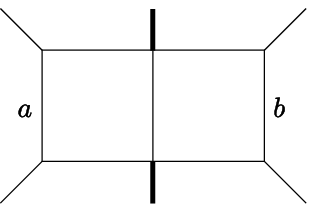}
\end{minipage}
\begin{minipage}[c]{0.7\textwidth}
\begin{equation}
  \label{eq:DBoomoom}
  0
\end{equation}
\end{minipage}

\begin{minipage}[c]{0.3\textwidth}
  \includegraphics{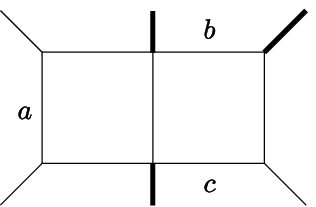}
\end{minipage}
\begin{minipage}[c]{0.7\textwidth}
\begin{equation}
  \label{eq:DBoommom}
  0
\end{equation}
\end{minipage}

\begin{minipage}[c]{0.3\textwidth}
  \includegraphics{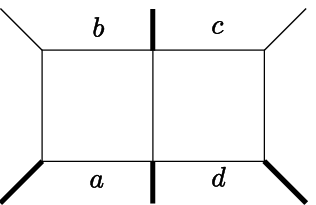}
\end{minipage}
\begin{minipage}[c]{0.7\textwidth}
\begin{equation}
  \label{eq:DBmomomm}
  0
\end{equation}
\end{minipage}

\begin{minipage}[c]{0.3\textwidth}
  \includegraphics{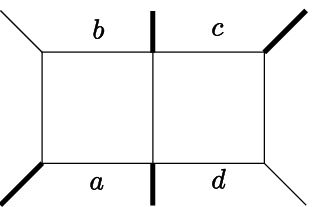}
\end{minipage}
\begin{minipage}[c]{0.7\textwidth}
\begin{equation}
  \label{eq:DBmommom}
  0
\end{equation}
\end{minipage}

\subsubsection{Extra double boxes}
\label{sssec:extra_double}

\begin{minipage}[c]{0.3\textwidth}
  \includegraphics{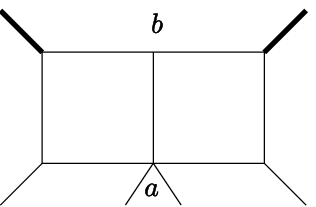}
\end{minipage}
\begin{minipage}[c]{0.7\textwidth}
\begin{multline}
  \label{eq:DBomzmot}
\frac{1}{4} \Bigl(-x_{a-2,b}^2 x_{a-1,a+2}^2 x_{a+1,b}^2+x_{a-2,a+2}^2
   x_{a-1,b}^2 x_{a+1,b}^2+\\+x_{a-2,b}^2 x_{a-1,a+1}^2 x_{a+2,b}^2-x_{a-2,a+1}^2
   x_{a-1,b}^2 x_{a+2,b}^2\Bigr)
\end{multline}
\end{minipage}

\begin{minipage}[c]{0.3\textwidth}
  \includegraphics{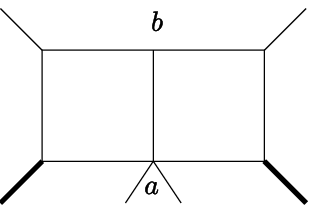}
\end{minipage}
\begin{minipage}[c]{0.7\textwidth}
\begin{equation}
  \label{eq:DBmozomt}
  -\frac{1}{4}
  \begin{bmatrix}
    a+1 & b-1 & b\\b & b+1 & a-1
  \end{bmatrix}
\end{equation}
\end{minipage}

\begin{minipage}[c]{0.3\textwidth}
  \includegraphics{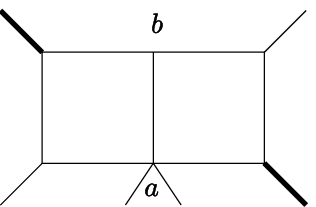}
\end{minipage}
\begin{minipage}[c]{0.7\textwidth}
\begin{equation}
  \label{eq:DBomzomt}
  0
\end{equation}
\end{minipage}

\begin{minipage}[c]{0.3\textwidth}
  \includegraphics{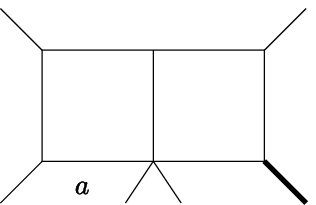}
\end{minipage}
\begin{minipage}[c]{0.7\textwidth}
\begin{equation}
  \label{eq:DBoozomt}
  -\frac{1}{4}
  \begin{bmatrix}
    a & a+1 & a+2\\a+2 & a+3 & a-2
  \end{bmatrix}
\end{equation}
\end{minipage}

\begin{minipage}[c]{0.3\textwidth}
  \includegraphics{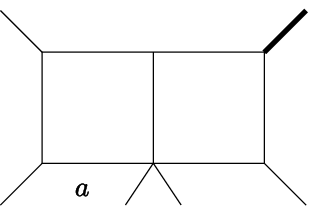}
\end{minipage}
\begin{minipage}[c]{0.7\textwidth}
\begin{equation}
  \label{eq:DBoozmot}
  \frac{1}{4} \left(x_{a-3,a+1}^2 x_{a-2,a+2}^2-x_{a-3,a+2}^2
   x_{a-2,a+1}^2\right) x_{a,a+2}^2
\end{equation}
\end{minipage}


\begin{minipage}[c]{0.3\textwidth}
  \includegraphics{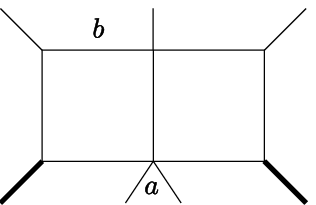}
\end{minipage}
\begin{minipage}[c]{0.7\textwidth}
\begin{equation}
  \label{eq:DBmooomt}
  -\frac{1}{4}
  \begin{bmatrix}
    a+1 & b-1 & b\\b+1 & b+2 & a-1
  \end{bmatrix}
\end{equation}
\end{minipage}

\begin{minipage}[c]{0.3\textwidth}
  \includegraphics{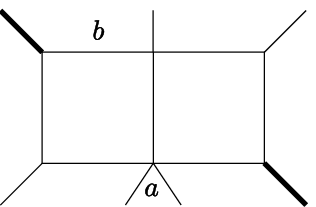}
\end{minipage}
\begin{minipage}[c]{0.7\textwidth}
\begin{equation}
  \label{eq:DBomoomt}
  0
\end{equation}
\end{minipage}

\begin{minipage}[c]{0.3\textwidth}
  \includegraphics{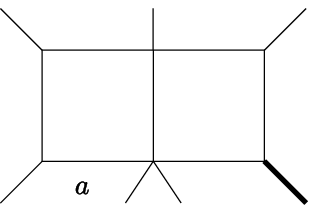}
\end{minipage}
\begin{minipage}[c]{0.7\textwidth}
\begin{equation}
  \label{eq:DBoooomt}
  -\frac{1}{4}
  \begin{bmatrix}
    a & a+1 & a+2\\a+3 & a+4 & a-2
  \end{bmatrix}
\end{equation}
\end{minipage}

\begin{minipage}[c]{0.3\textwidth}
  \includegraphics{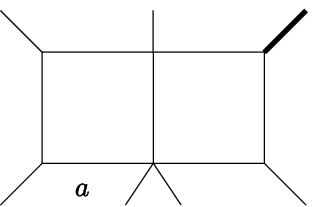}
\end{minipage}
\begin{minipage}[c]{0.7\textwidth}
\begin{equation}
  \label{eq:DBooomot}
\frac{1}{4} \left(x_{a-3,a+3}^2 x_{a-2,a+1}^2-x_{a-3,a+1}^2
   x_{a-2,a+3}^2\right) x_{a,a+2}^2
\end{equation}
\end{minipage}


\begin{minipage}[c]{0.3\textwidth}
  \includegraphics{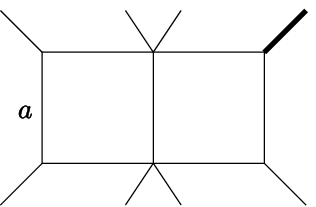}
\end{minipage}
\begin{minipage}[c]{0.7\textwidth}
\begin{equation}
  \label{eq:DBootmot}
  -\frac{1}{4}
  \begin{bmatrix}
    a-1 & a & a+1\\a+3 & a-4 & a-3
  \end{bmatrix}
\end{equation}
\end{minipage}

\begin{minipage}[c]{0.3\textwidth}
  \includegraphics{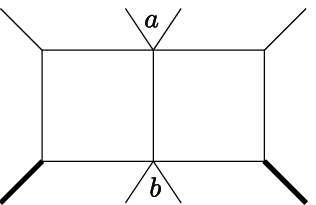}
\end{minipage}
\begin{minipage}[c]{0.7\textwidth}
\begin{equation}
  \label{eq:DBmotomt}
  0
\end{equation}
\end{minipage}

\begin{minipage}[c]{0.3\textwidth}
  \includegraphics{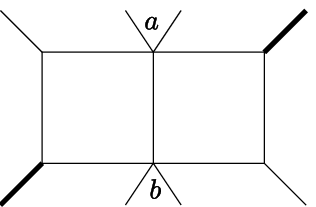}
\end{minipage}
\begin{minipage}[c]{0.7\textwidth}
\begin{equation}
  \label{eq:DBmotmot}
  0
\end{equation}
\end{minipage}

\begin{minipage}[c]{0.3\textwidth}
  \includegraphics{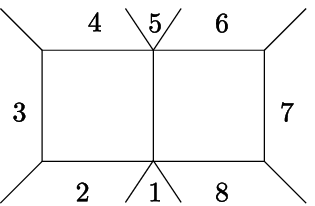}
\end{minipage}
\begin{minipage}[c]{0.7\textwidth}
\begin{equation}
  \label{eq:DBootoot}
  -\frac{1}{2}
  \begin{bmatrix}
    2 & 3 & 4\\6 & 7 & 8
  \end{bmatrix}
\end{equation}
\end{minipage}


\begin{minipage}[c]{0.3\textwidth}
  \includegraphics{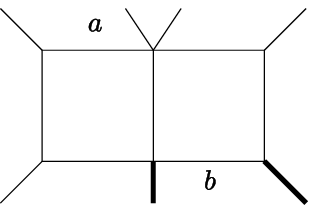}
\end{minipage}
\begin{minipage}[c]{0.7\textwidth}
\begin{equation}
  \label{eq:DBootomm}
  0
\end{equation}
\end{minipage}

\begin{minipage}[c]{0.3\textwidth}
  \includegraphics{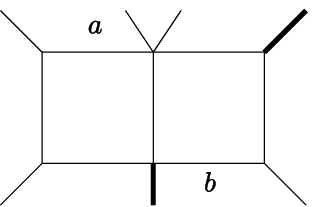}
\end{minipage}
\begin{minipage}[c]{0.7\textwidth}
\begin{equation}
  \label{eq:DBootmom}
  -\frac{1}{4}
  \begin{bmatrix}
    a-2 & a-1 & a\\a+2 & b-1 & b
  \end{bmatrix}
\end{equation}
\end{minipage}

\begin{minipage}[c]{0.3\textwidth}
  \includegraphics{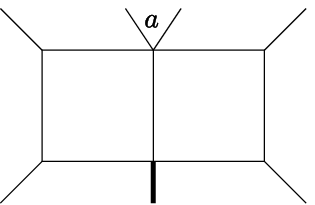}
\end{minipage}
\begin{minipage}[c]{0.7\textwidth}
\begin{equation}
  \label{eq:DBootoom}
  -\frac{1}{4}
  \begin{bmatrix}
    a-3 & a-2 & a-1\\a+1 & a+2 & a+3
  \end{bmatrix}
\end{equation}
\end{minipage}


\begin{minipage}[c]{0.3\textwidth}
  \includegraphics{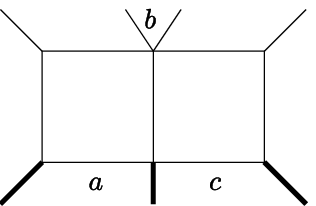}
\end{minipage}
\begin{minipage}[c]{0.7\textwidth}
\begin{equation}
  \label{eq:DBmotomm}
  0
\end{equation}
\end{minipage}

\begin{minipage}[c]{0.3\textwidth}
  \includegraphics{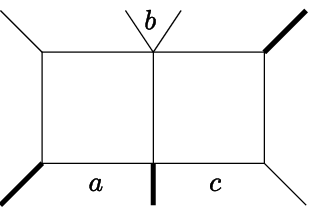}
\end{minipage}
\begin{minipage}[c]{0.7\textwidth}
\begin{equation}
  \label{eq:DBmotmom}
  0
\end{equation}
\end{minipage}

\begin{minipage}[c]{0.3\textwidth}
  \includegraphics{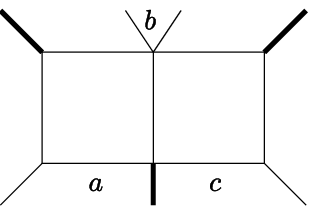}
\end{minipage}
\begin{minipage}[c]{0.7\textwidth}
\begin{equation}
  \label{eq:DBomtmom}
  -\frac{1}{4}
  \begin{bmatrix}
    a & a+1 & b-1\\b+1 & c-1 & c
  \end{bmatrix}
\end{equation}
\end{minipage}

\subsection{Kissing double-box topologies}

\begin{minipage}[t]{0.3\textwidth}
  \vspace{0pt}
  \includegraphics{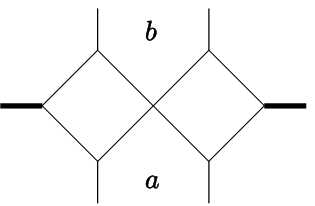}
\end{minipage}
\begin{minipage}[t]{0.7\textwidth}
\begin{multline}
  \label{eq:KBomozomoz}
  -\frac{1}{4}
  \begin{bmatrix}
    a & a+1 & b-1 & b\\ b & b+1 & a-1 & a
  \end{bmatrix} + \frac{1}{4}
  \begin{bmatrix}
    a & a+1\\b-1 & b
  \end{bmatrix}
  \begin{bmatrix}
    b & b+1\\a-1 & a
  \end{bmatrix} =\\
   \frac{1}{4} \Bigl(x_{a-1,b+1}^2 x_{a+1,b-1}^2
   \left(x_{a b}^2\right){}^2-x_{a-1,b-1}^2 x_{a+1,b+1}^2
   \left(x_{a b}^2\right){}^2+\\+x_{a-1,a+1}^2 x_{b-1,b+1}^2
   \left(x_{a b}^2\right){}^2-x_{a-1,b}^2 x_{a,b+1}^2 x_{a+1,b-1}^2
   x_{a b}^2-\\-x_{a-1,b+1}^2 x_{a,b-1}^2 x_{a+1,b}^2 x_{a b}^2+x_{a-1,b-1}^2
   x_{a,b+1}^2 x_{a+1,b}^2 x_{a b}^2+\\+x_{a-1,b}^2 x_{a,b-1}^2 x_{a+1,b+1}^2 x_{a b}^2\Bigr)
\end{multline}
\end{minipage}

\begin{minipage}[c]{0.3\textwidth}
  \includegraphics{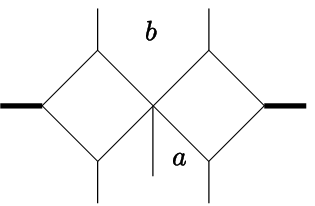}
\end{minipage}
\begin{minipage}[c]{0.7\textwidth}
\begin{equation}
  \label{eq:KBomozomoo}
  -\frac{1}{4}
  \begin{bmatrix}
    a+1 & a+2 & b-1 & b\\b & b+1 & a-1 & a
  \end{bmatrix} + \frac{1}{4}
  \begin{bmatrix}
    a-1 & a\\b & b+1
  \end{bmatrix}
  \begin{bmatrix}
    a+1 & a+2\\b-1 & b
  \end{bmatrix}
\end{equation}
\end{minipage}

\begin{minipage}[c]{0.3\textwidth}
  \includegraphics{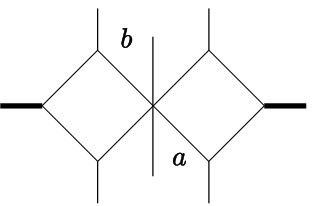}
\end{minipage}
\begin{minipage}[c]{0.7\textwidth}
\begin{equation}
  \label{eq:KBomooomoo}
  -\frac{1}{4}
  \begin{bmatrix}
    a+1 & a+2 & b-1 & b\\b+1 & b+2 & a-1 & a
  \end{bmatrix} + \frac{1}{4}
  \begin{bmatrix}
    a+1 & a+2\\b-1 & b
  \end{bmatrix}
  \begin{bmatrix}
    b+1 & b+2\\a-1 & a
  \end{bmatrix}
\end{equation}
\end{minipage}

\begin{minipage}[c]{0.3\textwidth}
  \includegraphics{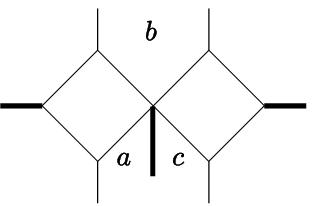}
\end{minipage}
\begin{minipage}[c]{0.7\textwidth}
\begin{equation}
  \label{eq:KBomozomom}
  -\frac{1}{4}
  \begin{bmatrix}
    a & a+1 & b-1 & b\\b & b+1 & c-1 & c
  \end{bmatrix} + \frac{1}{4}
  \begin{bmatrix}
    a & a+1\\b-1 & b
  \end{bmatrix}
  \begin{bmatrix}
    b & b+1\\c-1 & c
  \end{bmatrix}
\end{equation}
\end{minipage}

\begin{minipage}[c]{0.3\textwidth}
  \includegraphics{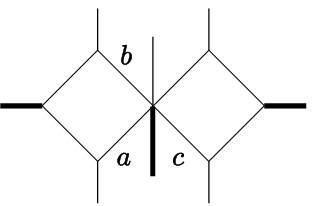}
\end{minipage}
\begin{minipage}[c]{0.7\textwidth}
\begin{equation}
  \label{eq:KBomooomom}
  -\frac{1}{4}
  \begin{bmatrix}
    a & a+1 & b-1 & b\\b+1 & b+2 & c-1 & c
  \end{bmatrix} + \frac{1}{4}
  \begin{bmatrix}
    a & a+1\\b-1 & b
  \end{bmatrix}
  \begin{bmatrix}
    b+1 & b+2\\c-1 & c
  \end{bmatrix}
\end{equation}
\end{minipage}

\begin{minipage}[c]{0.3\textwidth}
  \includegraphics{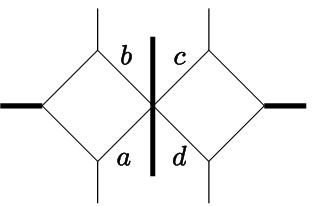}
\end{minipage}
\begin{minipage}[c]{0.7\textwidth}
\begin{equation}
  \label{eq:KBomomomom}
  -\frac{1}{4}
  \begin{bmatrix}
    a & a+1 & b-1 & b\\c & c+1 & d-1 & d
  \end{bmatrix} + \frac{1}{4}
  \begin{bmatrix}
    a & a+1\\b-1 & b
  \end{bmatrix}
  \begin{bmatrix}
    c & c+1\\d-1 & d
  \end{bmatrix}
\end{equation}
\end{minipage}

\subsection{Box-Pentagon topologies}

\subsubsection{No legs attached}

\begin{minipage}[c]{0.3\textwidth}
  \includegraphics{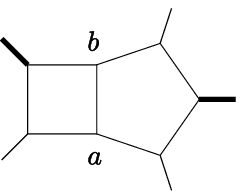}
\end{minipage}
\begin{minipage}[c]{0.7\textwidth}
\begin{equation}
  \label{eq:BPomzomoz}
\frac{1}{4} x_{a b}^2 x_{a+1,q}^2 \left(x_{a,b+1}^2 x_{a-1,b}^2-x_{a b}^2 x_{a-1,b+1}^2\right)
\end{equation}
\end{minipage}

\begin{minipage}[c]{0.3\textwidth}
  \includegraphics{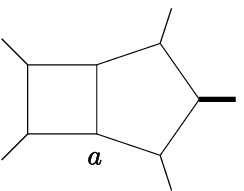}
\end{minipage}
\begin{minipage}[c]{0.7\textwidth}
\begin{equation}
  \label{eq:BPoozomoz}
\frac{1}{2} x_{a,a+2}^2 x_{a+1,q}^2 \left(x_{a-1,a+2}^2 x_{a,a+3}^2-x_{a-1,a+3}^2 x_{a,a+2}^2\right)
\end{equation}
\end{minipage}

\subsubsection{One massless leg attached}

\begin{minipage}[c]{0.3\textwidth}
  \includegraphics{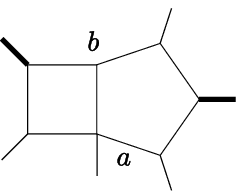}
\end{minipage}
\begin{minipage}[c]{0.7\textwidth}
\begin{equation}
  \label{eq:BPomzomoo}
\frac{1}{4} \left(x_{a-1,b+1}^2 x_{ab}^2-x_{a-1,b}^2 x_{a,b+1}^2\right)
   \left(x_{a+1,q}^2 x_{a+2,b}^2-x_{a+1,b}^2 x_{a+2,q}^2\right)
\end{equation}
\end{minipage}

\begin{minipage}[c]{0.3\textwidth}
  \includegraphics{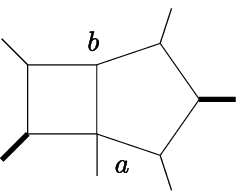}
\end{minipage}
\begin{minipage}[c]{0.7\textwidth}
\begin{equation}
  \label{eq:BPmozomoo}
  \frac{1}{4} x_{a-1,b}^2 \left(x_{ab}^2 x_{a+1,q}^2 x_{b-1,b+1}^2+x_{a,b+1}^2
   x_{a+1,b}^2 x_{b-1,q}^2-x_{ab}^2 x_{a+1,b+1}^2 x_{b-1,q}^2\right)
\end{equation}
\end{minipage}

\begin{minipage}[c]{0.3\textwidth}
  \includegraphics{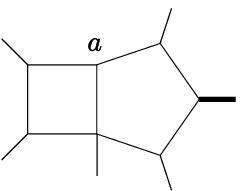}
\end{minipage}
\begin{minipage}[c]{0.7\textwidth}
\begin{multline}
  \label{eq:BPoozomoo}
  \frac{1}{4} \Bigl(x_{a-4,a}^2 x_{a-3,a}^2 x_{a-2,q}^2
   x_{a-1,a+1}^2-x_{a-4,a+1}^2 x_{a-3,a}^2 x_{a-2,a}^2 x_{a-1,q}^2+\\+2 x_{a-4,a}^2 x_{a-3,a+1}^2 x_{a-2,a}^2 x_{a-1,q}^2-x_{a-4,a}^2 x_{a-3,a}^2 x_{a-2,a+1}^2 x_{a-1,q}^2\Bigr)
\end{multline}
\end{minipage}

\subsubsection{One massive leg attached}

\begin{minipage}[c]{0.3\textwidth}
  \includegraphics{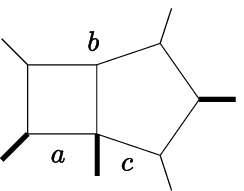}
\end{minipage}
\begin{minipage}[c]{0.7\textwidth}
\begin{equation}
  \label{eq:BPmozomom}
  0
\end{equation}
\end{minipage}

\begin{minipage}[c]{0.3\textwidth}
  \includegraphics{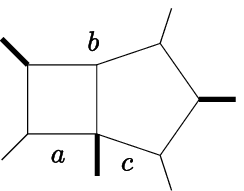}
\end{minipage}
\begin{minipage}[c]{0.7\textwidth}
\begin{equation}
  \label{eq:BPomzomom}
  \frac{1}{4} \left(x_{aq}^2 x_{a+1,b}^2-x_{ab}^2 x_{a+1,q}^2\right)
   \left(x_{bc}^2 x_{b+1,c-1}^2-x_{b,c-1}^2 x_{b+1,c}^2\right)
\end{equation}
\end{minipage}

\begin{minipage}[c]{0.3\textwidth}
  \includegraphics{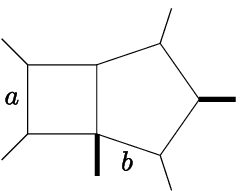}
\end{minipage}
\begin{minipage}[c]{0.7\textwidth}
\begin{equation}
  \label{eq:BPoozomom}
  \frac{1}{4} x_{a-1,a+1}^2 x_{aq}^2 \left(x_{a+1,b-1}^2 x_{a+2,b}^2-x_{a+1,b}^2 x_{a+2,b-1}^2\right)
\end{equation}
\end{minipage}

\subsubsection{One massless, one massive leg attached}

\begin{minipage}[c]{0.3\textwidth}
  \includegraphics{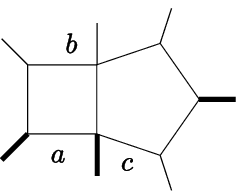}
\end{minipage}
\begin{minipage}[c]{0.7\textwidth}
\begin{equation}
  \label{eq:BPmooomom}
  0
\end{equation}
\end{minipage}

\begin{minipage}[c]{0.3\textwidth}
  \includegraphics{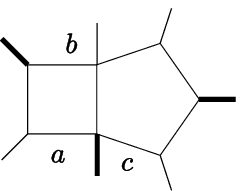}
\end{minipage}
\begin{minipage}[c]{0.7\textwidth}
\begin{equation}
  \label{eq:BPomoomom}
  -\frac{1}{4}
  \begin{bmatrix}
    a & a+1 & b & b+1\\b+2 & c-1 & c & q
  \end{bmatrix}.
\end{equation}
\end{minipage}

Note that in the previous formula we suppress the terms containing $x_{b+1,q}^2$ which would otherwise cancel a propagator of the underlying topology.  When expanded out, the expression above has 12 terms.

\begin{minipage}[c]{0.3\textwidth}
  \includegraphics{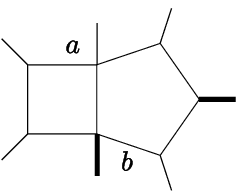}
\end{minipage}
\begin{minipage}[c]{0.7\textwidth}
\begin{equation}
  \label{eq:BPoooomom}
  -\frac{1}{4}
  \begin{bmatrix}
    a-2 & a-1 & a & a+1\\a+2 & b-1 & b & q
  \end{bmatrix}.
\end{equation}
\end{minipage}

In the previous formula we suppress the terms containing $x_{a+1,q}^2$ which would otherwise cancel a propagator of the underlying topology.

\subsubsection{Two massless legs attached}

\begin{minipage}[c]{0.3\textwidth}
  \includegraphics{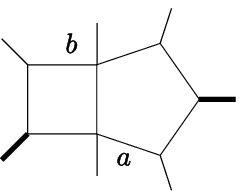}
\end{minipage}
\begin{minipage}[c]{0.7\textwidth}
\begin{equation}
  \label{eq:BPmooomoo}
  \frac{1}{4}
  \begin{bmatrix}
    a & a+1 & b-1 & b\\b+1 & b+2 & a-1 & q
  \end{bmatrix}
\end{equation}
\end{minipage}

In the previous formula we suppress the terms containing $x_{a+1,q}^2$ which would otherwise cancel a propagator of the underlying topology.

\begin{minipage}[t]{0.3\textwidth}
  \vspace{0pt}
  \includegraphics{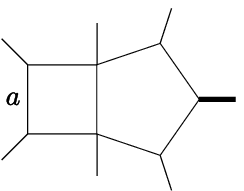}
\end{minipage}
\begin{minipage}[t]{0.7\textwidth}
\begin{multline}
  \label{eq:BPoooomoo}
  \frac{1}{4}
  \begin{bmatrix}
    a-2 & a-1 & a & a+1\\a+2 & a+3 & a-3 & q
  \end{bmatrix} -\frac{1}{4}
  \begin{bmatrix}
    a-1 & a & a+1 & a+2\\a+3 & a-3 & a-2 & q
  \end{bmatrix} =\\
  \frac{1}{4}
\Bigl(-x_{a-3,a+1}^2 x_{a-2,a+3}^2 x_{a-1,q}^2 x_{a,a+2}^2+
x_{a-3,a-1}^2 x_{a-2,a+3}^2 x_{a+1,q}^2 x_{a,a+2}^2-\\-
x_{a-3,a+2}^2 x_{a-2,a+1}^2 x_{a-1,q}^2 x_{a,a+3}^2+
2 x_{a-3,a+1}^2 x_{a-2,a+2}^2 x_{a-1,q}^2 x_{a,a+3}^2+\\+
x_{a-3,a+1}^2 x_{a-2,a+3}^2 x_{a-1,a+2}^2 x_{a q}^2+
x_{a-3,a+2}^2 x_{a-2,a+1}^2 x_{a-1,a+3}^2 x_{a q}^2-\\-
2 x_{a-3,a+1}^2 x_{a-2,a+2}^2 x_{a-1,a+3}^2 x_{a q}^2+
x_{a-3,a+2}^2 x_{a-2,a}^2 x_{a-1,q}^2 x_{a+1,a+3}^2-\\-
2 x_{a-3,a}^2 x_{a-2,a+2}^2 x_{a-1,q}^2 x_{a+1,a+3}^2+
2 x_{a-3,a-1}^2 x_{a-2,a+2}^2 x_{a q}^2 x_{a+1,a+3}^2-\\-
x_{a-3,a}^2 x_{a-2,a+3}^2 x_{a-1,a+2}^2 x_{a+1,q}^2-
x_{a-3,a+2}^2 x_{a-2,a}^2 x_{a-1,a+3}^2 x_{a+1,q}^2+\\+
2 x_{a-3,a}^2 x_{a-2,a+2}^2 x_{a-1,a+3}^2 x_{a+1,q}^2-
2 x_{a-3,a-1}^2 x_{a-2,a+2}^2 x_{a,a+3}^2 x_{a+1,q}^2\Bigr).
\end{multline}
\end{minipage}

We have written down this formula to emphasize how nontrivial it is.  We suppress the terms containing $x_{a-2,q}^2$ and $x_{a+2,q}^2$, respectively.  These terms would otherwise cancel a propagator of the underlying topology.
We will see below that the box-pentagon topologies with massless legs attached to the vertices of the edge common to both loops can in fact be seen to originate in double-pentagon topologies, by cancelling some propagators.

\subsection{Double pentagon topologies}

\subsubsection{No legs attached}

\begin{minipage}[c]{0.3\textwidth}
  \includegraphics{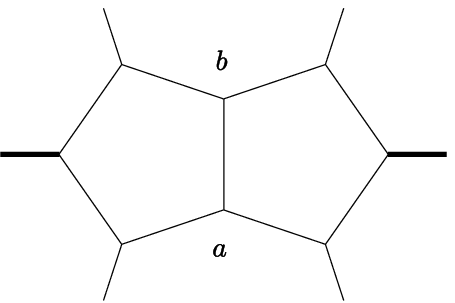}
\end{minipage}
\begin{minipage}[c]{0.7\textwidth}
\begin{equation}
  \label{eq:DPomozomoz}
  -\frac{1}{4}
  \begin{bmatrix}
    a & a+1 & b-1 & b & p\\b & b+1 & a-1 & a & q
  \end{bmatrix}
\end{equation}
\end{minipage}

In the expansion of the above formula we drop terms that would cancel propagators (in this case, the terms containing $x_{a p}^2$, $x_{a q}^2$, $x_{b p}^2$, $x_{b q}^2$, or $x_{p q}^2$).  This expression has $6$ terms when expanded.

\subsubsection{One massless leg attached}

\begin{minipage}[c]{0.3\textwidth}
  \includegraphics{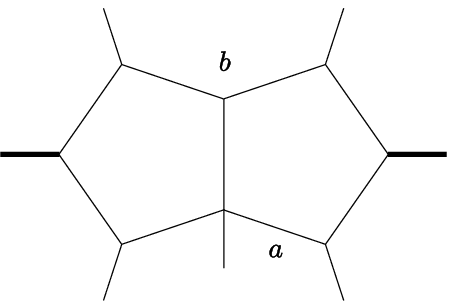}
\end{minipage}
\begin{minipage}[c]{0.7\textwidth}
\begin{equation}
  \label{eq:DPomozomoo}
  -\frac{1}{4}
  \begin{bmatrix}
    a+1 & a+2 & b-1 & b & p\\b & b+1 & a-1 & a & q
  \end{bmatrix}
\end{equation}
\end{minipage}

In the formula above we drop terms that would cancel propagators (in this case, the terms are $x_{b p}^2$, $x_{b q}^2$ and $x_{p q}^2$).  This expression has $15$ terms when expanded.

\subsubsection{One massive leg attached}

\begin{minipage}[c]{0.3\textwidth}
  \includegraphics{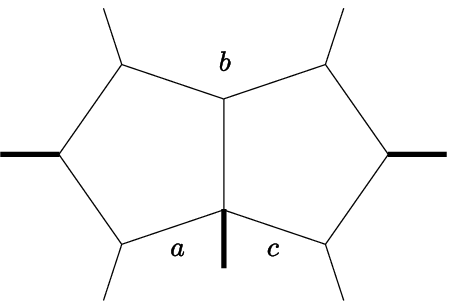}
\end{minipage}
\begin{minipage}[c]{0.7\textwidth}
\begin{equation}
  \label{eq:DPomozomom}
  -\frac{1}{4}
  \begin{bmatrix}
    a & a+1 & b-1 & b & p\\b & b+1 & c-1 & c & q
  \end{bmatrix}
\end{equation}
\end{minipage}

In the formula above we drop terms that would cancel propagators (in this case, the terms containing $x_{b p}^2$, $x_{b q}^2$ or $x_{p q}^2$).  This expression has $16$ terms when expanded.

\subsubsection{Two massless legs attached}

\begin{minipage}[c]{0.3\textwidth}
  \includegraphics{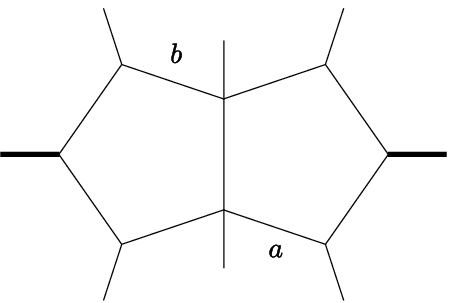}
\end{minipage}
\begin{minipage}[c]{0.7\textwidth}
\begin{equation}
  \label{eq:DPomooomoo}
  -\frac{1}{4}
  \begin{bmatrix}
    a+1 & a+2 & b-1 & b & p\\b+1 & b+2 & a-1 & a & q
  \end{bmatrix}
\end{equation}
\end{minipage}

In the formula we drop terms that would cancel propagators (in this case, the terms containing $x_{p q}^2$).  This expression has $64$ terms when expanded.

\subsubsection{One massless, one massive leg attached}

\begin{minipage}[c]{0.3\textwidth}
  \includegraphics{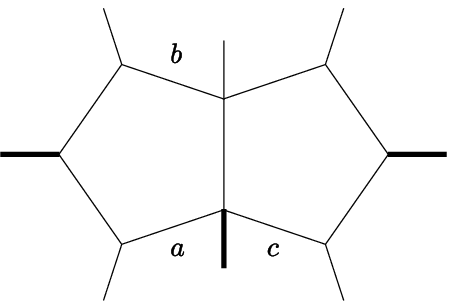}
\end{minipage}
\begin{minipage}[c]{0.7\textwidth}
\begin{equation}
  \label{eq:DPomooomom}
  -\frac{1}{4}
  \begin{bmatrix}
    a & a+1 & b-1 & b & p\\b+1 & b+2 & c-1 & c & q
  \end{bmatrix}
\end{equation}
\end{minipage}

In the formula above we drop terms that would cancel propagators (in this case, the terms containing $x_{p q}^2$).  This expression has $78$ terms when expanded.

\subsubsection{Two massive legs attached}

\begin{minipage}[c]{0.3\textwidth}
  \includegraphics{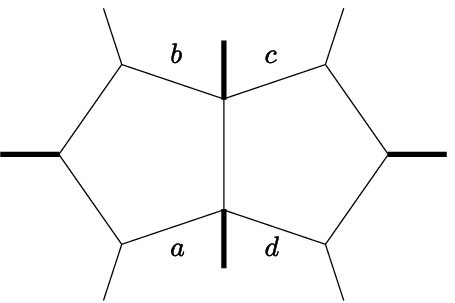}
\end{minipage}
\begin{minipage}[c]{0.7\textwidth}
\begin{equation}
  \label{eq:DPomomomom}
  -\frac{1}{4}
  \begin{bmatrix}
    a & a+1 & b-1 & b & p\\c & c+1 & d-1 & d & q
  \end{bmatrix}
\end{equation}
\end{minipage}

In the formula above we drop terms that would cancel propagators (in this case, the terms containing $x_{p q}^2$).  When expanded, the above expression contains $96$ terms.  The number of conformal dressings is $160$ (the number of coefficients unrelated by symmetries is lower).

\subsection{Assembly of the result}

As explained in Sec.~\ref{sec:review}, for the MHV amplitudes the ratio between the $\ell$-loop amplitude and the tree-level amplitude can be written as a sum between parity even and parity odd contributions
\begin{equation}
  M_n^{(\ell)} = M_n^{(\ell), \text{even}} + M_n^{(\ell), \text{odd}}.
\end{equation}  Then, the even part can be written
\begin{equation}
  \label{eq:even}
  M_n^{(2), \text{even}} = - \pi^{-D} e^{2 \gamma \epsilon} \int d^D x_p d^D x_q \sum_\sigma \sum_{i \in \text{Topologies}} s_i c_i I_i,
\end{equation} where the first sum runs over cyclic and anti-cyclic permutations of the external legs, the second sum runs over all the topologies, $s_i$ is a symmetry factor associated to topology $i$, $c_i$ is the numerator of the topology $i$, as listed in Sec.~\ref{sec:results} and $I_i$ is the denominator or the product of propagators in the topology $i$.

Apart from the parity odd part which we have not computed, there is also a contribution which is not detectable from four-dimensional cuts, denoted by $M^{(2), \mu}$.  This part of the result is such that its integrand vanishes in four dimensions, but the integral itself can give contributions to the divergent and finite parts.  In Ref.~\cite{Bern:2008ap}, for $n=6$ case, this part of the result was found to be closely related to $\mathcal{O}(\epsilon)$ contributions at one loop, $M^{(1),\mu}$.

Based on previous computations we expect that the odd part and the $\mu$ integrals will not be needed in order to compare with the Wilson loop results.  The odd parts could be computed by using the leading singularity method (see Ref.~\cite{Cachazo:2008vp} and also~\cite{Cachazo:2008dx, Buchbinder:2005wp}) or the technique of maximal cuts of Ref.~\cite{Bern:2007ct}.  In order to compute $M^{(2), \mu}$, one would have to compute $D$-dimensional cuts.  In practice this is done by computing the cuts of $\mathcal{N}=1$ super-Yang-Mills in ten dimensions, dimensionally reduced to $D$ dimensions.

\section{Discussion}

In this paper we computed the even part of the two-loop planar MHV scattering amplitudes in $\mathcal{N}=4$ super Yang-Mills.  The answer can be expressed in terms of a finite (and relatively small) number of two-loop pseudo-conformal integrals.

A computation of these integrals in dimensional regularization through the finite parts (of order $\mathcal{O}(\epsilon^0)$) would be very interesting and would allow a comparison with the results of Ref.~\cite{Anastasiou:2009kn}, where the corresponding Wilson loop computation was performed.

However, a computation of these integrals seems to be rather difficult.  In Ref.~\cite{Anastasiou:2009kn} the Wilson loop result was expressed in terms of some master integrals called: ``hard'', ``curtain'', ``cross'', ``Y'' and ``factorized cross.''  These master integrals depend on whether some momenta are zero, massless or massive (this is similar to the situation for scattering amplitudes; in that case also, the value of the integral depends on whether the external legs are massive or massless).

It is interesting to note that for the Wilson loop computation, there are no new master integrals beyond nine sides (this number arises by considering the ``hard'' integral where all the momenta $Q_1$, $Q_2$ and $Q_3$ are massive).  For the scattering amplitude, however, new integrals appear until twelve points, as shown in this paper.  It would be interesting to get a deeper understanding of this ``mismatch.''

The results presented in this paper hint that a different organization of the result may be possible.  For example, the coefficients written down using the square brackets symbols can be assembled over a common denominator whose topology is that of a double pentagon.  Sometimes, the coefficient of a given topology needs to be split into two contributions which get assembled into different double pentagon topologies (see Eq.~\eqref{eq:BPoooomoo} for an example).

It is also noteworthy that part of the kissing double boxes coefficient neatly combines with a double pentagon topology after multiplying the numerator and denominator by $x_{p q}^2$, while the remaining part has a factorized form.  This factorized form is a product of ``one-mass'' and ``two-mass easy'' integrals which appear in the one-loop MHV computation, but it is not equal to the square of the one-loop amplitude.

\acknowledgments

I would like to thank the Mathematical Institute in Oxford and especially David Skinner and Lionel Mason for hospitality.  I am also grateful to the Stony Brook University and the Simons Center for Geometry and Physics for an enjoyable week spent at the Simons Workshop 2009.

It is a pleasure to thank Paul Heslop and Gabriele Travaglini for discussions about the Wilson loop computations and to Marcus Spradlin and Anastasia Volovich for discussions and for their kind interest in this work.

This research is supported in part by the US Department of Energy under contract DE-FG02-91ER40688 and by the US National Science Foundation under grant PHY-0643150.

\end{document}